\documentclass[a4paper,12pt]{article}
\usepackage{amsmath,amsthm,amssymb,bbm}

\usepackage[english]{babel}
\usepackage{graphicx}

\newcommand{\ud}{\mathrm{d}}				
\newcommand{\ket}[1]{\bigl|#1\bigr\rangle}			% per i ket
\newcommand{\bra}[1]{\bigl\langle#1\bigr|}			% per i bra
\newcommand{\comm}[2]{\left[#1 \ , \ #2\right]} % per i commutatori
\newcommand{\proj}[2]{\ket{#1}\bra{#2}}				% per i proiettori
\newcommand{\scalarprod}[2]{\bigl\langle#1\big| #2\bigr\rangle}	% per i prodotti scalari tra vettori
\newcommand{\bi}{\hat{b}}					% per l'operatore bi (senza indice)
\newcommand{\bid}{\hat{b}^{\dagger}}				% per l'operatore bi daga (senza indice)
\title{\bf Quantum measures for density correlations\\ in optical lattices}

\author{F. Benatti$^{a,b}$, 
R. Floreanini$^{b}$ and G. G. Guerreschi$^{a}$\\
\small ${}^a$Dipartimento di Fisica Teorica, Universit\`a di Trieste, 
34014 Trieste, Italy\\
\small ${}^b$Istituto Nazionale di Fisica Nucleare, Sezione di Trieste,
34014 Trieste, Italy}

\date{\null}

\begin{document}

\maketitle

\begin{abstract}
\noindent
The density-density correlation profiles obtained superimposing absorption images from atomic clouds
freely expanding after the release of the confining optical lattice can be theoretically described
in terms of a generalized quantum measure based on coherent-like states. 
We show that the corresponding density patterns differ in a testable way from 
those computed using standard many-body mean values, usually adopted in fitting experimental data.
\end{abstract}

\section{Introduction}

A standard technique used in experiments for extracting information on the behavior
of ultracold atomic gases trapped in optical lattices%
\footnote{For recent reviews on the study of quantum many-body effects in these systems,
see \cite{Lewenstein,Bloch}.}
is based on analysis 
of interference phenomena (see \cite{Stringari}-\cite{Demler} and references therein). 
Since the measure of relevant observables inside the optical lattice
is problematic, the usual adopted procedure consists in the release of the confining optical potentials,
followed by the (free) expansion of the atomic cloud up to mesoscopic sizes.
At this point, the cloud is illuminated by a laser beam and the corresponding absorption image
collected. The absorption process is destructive; nevertheless, many pictures can be
obtained by starting each time with a new system, prepared in the same initial state.
By superimposing the various obtained pictures, 
information on the atom density at the moment of trap release can be inferred.

The image that is obtained by superimposing all these ``photographs'' is usually interpreted as the average 
of the density operator over the state of the sample. 
An alternate theoretical description is however possible. Indeed, in quantum mechanics, any averaging
procedure obtained through a measuring process corresponds to a generalized
quantum measure, {\it i.e.} a so-called {\it Positive Operator Valued Measure} (POVM) \cite{Takesaki}-\cite{Alicki}. 
In general, the choice of the POVM to be used is suggested by the experimental 
evidences. In the present case of ultracold atoms in optical lattices,
this evidence comes from experiments involving
a two-well trapping potential \cite{Andrews}. 
The actual data show that interference effects appear in a single absorption image even when the 
system is prepared in a totally incoherent state: more precisely, the interference pattern seen in
single pictures seems to always conform to what is expected for a condensed, fixed-phase state, a state in which
all atoms share the same single particle wave function. By constructing a POVM in terms of
these coherent-like states, one finds that the corresponding 
generalized quantum measurement process leads to predictions
that, at least in line of principle, differ from those obtained 
through the simple average of the density operator \cite{ABFT}.%
\footnote{The standard theoretical intepretation of these experimental facts explains the
appearence of the density interference fringes as an ``emergent phenomena''\cite{Javanainen}-\cite{Castin};
although phenomenological in character, also this averaging procedure can be interpreted
in terms of a quantum generalized measure, albeit through a rather involved,
unnatural POVM.}

The aim of the present investigation is to analyze a possible physical scenario in which
those differences may become visible and experimentally detectable. 
To this end, we shall study the behavior of a system of cold atoms in bichromatic optical lattices, where a second,
low-intensity laser is superimposed to the one forming the periodic potential with the aim of obtaining
an unbalanced filling of the lattice sites \cite{Fallani1}-\cite{Fallani4}. 
By exploiting the properties of the density-density correlation
function \cite{Bloch2}, one can show that there are experimentally 
relevant instances in which the differences in the predictions
of the two above mentioned theoretical interpretations can be revealed. This result has been supported by
a numerical simulation, reproducing the situation of an actual experimental setup.
We are confident that these results will stimulate further direct analysis and tests.

\section{Cold bosonic gases in optical lattices}

We shall study the behavior of $N$ bosons confined in one-dimensional lattice with $M$ sites,
each separated by a fixed distance $d$.%
\footnote{In the actual experimental setups, the bosons are really confined 
in an three-dimensional harmonic trap over which a periodic potential along say the $x$ direction is superimposed.
Since the confining potentials are separable, 
the single particle wave function describing the state of the
atoms factorizes in a part depending only on the variable $x$ times a piece depending on the couple $(y,z)$.
The dependence on the transverse coordinates is irrelevant for the dynamics in the lattice, that therefore
can be effectively described by a one-dimensional Hamiltonian.}
In a suitable approximation, {\it i.e.} for a large enough inter-site barriers, their dynamics can be
described by a Bose-Hubbard Hamiltonian \cite{Lewenstein,Bloch},
\begin{equation}
\label{BH}
H = -J \sum_{<i,j>} \bid_{i}\, \bi_{j} + \sum _{i} \epsilon_{i}\, \hat{n}_{i} + \frac{1}{2}U \sum_{i} \hat{n}_{i}\left( \hat{n}_{i} -1 \right) \ ,\quad i,j=1,2,\ldots,M\ ,
\end{equation}
where $<i,j>$ means nearest neighbor, $\hat{b}_i^\dagger$, $\hat{b}_i$ are creation and annihilation operator
for an atom in site $i$, $\hat{n}_{i}=\bid_{i}\bi_{i}$ is the number operator on site $i$, while
$J$, $U$ and $\epsilon_i$ are parameters that quantify the hopping, repulsion and single-site depth energy, 
respectively. The operators $\bid_{i}$, $\bi_{i}$ obey the standard Bose commutation relations:
$[\bi_{i}, \bid_{j}]=\delta_{ij}$.

The total number of particles $N$ is conserved by the dynamics generated by (\ref{BH}). Therefore,
the Hilbert space of the system is ${N+M-1}\choose{N}$-dimensional 
and can be be spanned by the set of Fock states,
$|\vec{k};N\rangle$, describing the situation in which the occupation number of each site is fixed;
the $M$-dimensional vector $\vec{k}=(k_1,k_2,\ldots,k_M)$, with $\sum_{i=1}^M k_i=N$, 
represents a possible distributions of the $N$ atoms in the $M$ sites. 
These states are obtained by acting with the creation operators on the vacuum $\ket{0}$; explicitly, one has
\begin{equation}
\label{Fock}
\ket{\vec{k};N} \equiv \dfrac{1}{\sqrt{\vec{k}!}} 
(\bid_{1})^{k_{1}} (\bid_{2})^{k_{2}}\ldots(\bid_{M})^{k_{M}} \ket{0} \ ,
\end{equation}
where in short $\vec{k}!=k_1!\ k_2!\ldots k_{M}!$.

A different basis in the system Hilbert space is given by the collection of fixed-phase, coherent-like states,
\begin{equation}
\label{coherent}
\ket{\vec{\xi},\vec{\varphi};N} \equiv \dfrac{1}{\sqrt{N!}}\Bigg( \sum_{j=1}^{M} e^{i\varphi_{j}}\sqrt{\xi_{j}}\ \bid_{j} \Bigg) ^{N} \ket{0} \ ,
\end{equation}
where $\vec{\xi},\vec{\varphi}$ are $M$-dimensional vectors whose components $\xi_j,\varphi_j$, $j=1,2,...,M$,
represent sets of real parameters such that $\varphi_j\in[0,2\pi]$, $\xi_j\in[0,1]$ with $\sum_{j}\xi_{j}=1$.
These states describe a physical situation in which all $N$ atoms are in a coherent superposition,
where $\xi_i$ measures the probability of finding an atom in the $i$-th site, while $\varphi_i$ gives the
corresponding phase. Only relative phases are relevant, so that one can arbitrarily fix the value of one of the
$\varphi_i$. 

The Fock states form an orthonormal set,
while the coherent ones become orthogonal only in the large $N$ limit;%
\footnote{
Indeed, for an $M$-site lattice,
one finds that: 
$\langle\vec\xi,\vec\varphi;N | \vec\xi',\vec\varphi'; N\rangle=
\Big(\sum_{i=1}^M \sqrt{\xi_i \xi'_i} e^{i(\varphi_i-\varphi'_i)}\Big)^N$;
using the Cauchy-Schwartz inequality, one easily sees that
the modulus of sum in the bracket is always less than one,
unless $\vec\xi=\vec\xi'$ and $\vec\varphi=\vec\varphi'$; as a consequence,
its $N$-th power become vanishingly small as $N$ become large.}
nevertheless, they form an overcomplete set of states \cite{ABFT}:
\begin{equation}
\label{overcomplete}
\mathbbm{1}=\dfrac{(N+M-1)!}{N!} \int_{0}^{2\pi}\frac{d\varphi_{1}}{2\pi}\ldots\frac{{\ud\varphi_{M-1}}}{2\pi}
\int_{0}^{1} \ud\xi_{1}\ldots\int_{1}^{1-\xi_{1}-\xi_{2}...-\xi_{M-2}} \ud\xi_{M-1}\ 
\proj{\vec{\xi},\vec{\varphi};N}{\vec{\xi},\vec{\varphi};N} \ .
\end{equation}
By expanding a coherent state over the Fock basis, one finds:
\begin{equation}
\label{comodo1}
\ket{\vec{\xi};\vec{\varphi};N} = \sum_{\vec{k}} \sqrt{\dfrac{N!}{\vec{k}!}} e^{i\vec{k}\cdot\vec{\varphi}} \Bigg( \prod_{j=1}^{M} (\xi_{j})^{\frac{k_{j}}{2}}\Bigg) \ket{\vec{k};N} \ ,
\end{equation}
where the sum runs over all possible $M$-vectors $\vec{k}$,
whose components $k_i$ obey the constraint $\sum_{i} k_{i}=N$. The overlap between a Fock and a coherent state
is then given by:
\begin{equation}
\label{fockfase}
\scalarprod{\vec{k};N}{\vec{\xi},\vec{\varphi};N} = \sqrt{\dfrac{N!}{\vec{k}!}} e^{i\vec{k}\cdot\vec{\varphi}} \Bigg( \prod_{j=1}^{M} (\xi_{j})^{\frac{k_{j}}{2}}\Bigg) \ .
\end{equation}

As well known \cite{Lewenstein,Bloch}, the Hamiltonian (\ref{BH}) describes a cross-over
between a superfluid and insulator phases, which becomes a true quantum phase transition,
with order parameter depending on the ratio $J/U$, in the limit of an infinite number $M$ of wells. 
Neglecting the shifts $\epsilon_i$,
for small $J/U$, the ground state of the system is given by
a Fock state (Mott insulator phase). On the other hand, for large $J/U$, 
the system shows phase coherence; all $N$ particles are in
the same superposition and the ground state of (\ref{BH}) can be approximated by
$\ket{\vec{\xi};\vec{\varphi};N}$, with definite relative phases and occupation probabilities (superfluid phase).

\section{Many-body states}

In a typical experimental setup, the system of $N$ atoms is first cooled to very low temperatures, of the order
of few tens of nanokelvin, and then trapped in the optical lattice. 
Since measures of relevant observables directly in the lattice are difficult, indirect information
on the dynamics of the atoms are usually obtained by switching off the periodic confining potential
and letting the atom gas expand freely up to mesoscopic dimensions. Absorption images of the expanded sample
are then collected by shining it with a probe laser; by superimposing the various obtained pictures, 
information on the atom density at the moment of trap release can be inferred.

In order to theoretically describe this process of measure, it is convenient to use
a second quantized many-body formalism.
Let us first introduce the field operator $\hat\psi^\dagger(x)$, creating from the vacuum an atom at position $x$,
$\hat\psi^\dagger(x)\, |0\rangle=|x\rangle$; it can be decomposed as
\begin{equation}
\label{psi}
\hat\psi^\dagger(x)=\sum_{i=1}^\infty \bar w_i(x)\, b^\dagger_i\ ,
\end{equation}
in terms of a complete set
of single-particle wave functions $w_i(x)\equiv \langle x|w_i\rangle=\langle x|\hat b_i^\dagger|0\rangle$.
Although the states $|w_i\rangle\equiv \hat b^\dagger_i |0\rangle$, $i=1,2\ldots,M$, obtained by the action
of the creation operators $\hat b_i^\dagger$ on the vacuum are enough to describe 
an atom confined in the lattice, a complete, infinite set is needed to properly
represent its state when it moves freely in space. These states are orthonormal and therefore one can invert
(\ref{psi}) and write
\begin{equation}
\bid_i=\int \ud x \ w_i(x) \hat{\psi}^\dagger(x) \ ,
\end{equation}
so that from $[\hat b_i,\hat b_j^\dagger]=\langle w_i|w_j\rangle=\delta_{ij}$ one recovers the standard
bosonic (equal-time) commutation relation, $\comm{\hat{\psi}(x)}{\hat{\psi}^{\dagger}(x')}=\delta (x-x')$.

Assume now that the confining potential is released at time $t=\,0$ and denote by
$\hat U_t$ the unitary operator that evolves freely in time the initial one-particle states:
\begin{equation}
\label{evolution}
\ket{w_i,t}\equiv \hat{U}_t \ket{w_i} = \hat{U}_t\, \bid_i \ket{0} = \bid_i (t) \ket{0} \ ,
\quad \bid_i(t) \equiv \hat{U}_t\, \bid_i\, \hat{U}_t^{\dagger}\ .
\end{equation}
The corresponding wave function is given by
$w_i(x;t)\equiv\langle x|w_i(t)\rangle = \langle x |\hat b^\dagger_i(t)|0\rangle$, and coincides
with a transformed $w_i(x)$ under a ballistic expansion.
At the moment of the release of the lattice, the $w_i(x,0)$, $i=1,2,\ldots, M$, 
are wave functions localized at the lattice sites $x_i$:
they can be identified with one-dimensional Wannier functions. After a large enough free expansion time $t$,
sufficient for the clouds coming from the various sites to overlap, 
one finds through (\ref{evolution}) that these functions have a common envelope, 
differing only by a phase \cite{Stringari,Pethick,Leggett}:
\begin{equation}
\label{evolw}
w_{i}(x,t)=|w(x,t)| e^{i\frac{m}{2\hbar t}(x-x_{i})^{2}} \ ;
\end{equation}
indeed, for those times, the scale over which $|w_i(x,t)|$ 
varies is larger than the product $Md$, giving the original dimension of the lattice.

Since the dynamics is free, every particle in a many-body state will evolve independently with $\hat U_t$;
therefore, the evolution up to time $t$ of the $t=\,0$ coherent state $\ket{\vec{\xi},\vec{\varphi};N}$
will simply be given by
\begin{equation}
\label{coherent-t}
\ket{\vec{\xi},\vec{\varphi};N,t} \equiv \dfrac{1}{\sqrt{N!}}\Bigg( \sum_{j=1}^{M} e^{i\varphi_{j}}\sqrt{\xi_{j}}\ \bid_{j}(t) \Bigg) ^{N} \ket{0} \ ,
\end{equation}
and analogously for Fock states (\ref{Fock})
\begin{equation}
\label{Fock-t}
\ket{\vec{k};N,t} = \dfrac{1}{\sqrt{\vec{k}!}} (\bid_{1}(t))^{k_{1}} (\bid_{2}(t))^{k_{2}}
\ldots(\bid_{M}(t))^{k_{M}} \ket{0} \ .
\end{equation}

In this picture, only states evolve in time while operators, like $\hat\psi(x)$, remain fixed. 
Since at each instant of time $t$ the collection $\{|w_i(t)\rangle\}_{i=1}^\infty$ 
is a complete set of single particle states obtained from the vacuum by the action of the creation operators
$\hat b^\dagger_i(t)$, $\hat\psi(x)$ can be equivalently decomposed as
$\hat\psi(x)=\sum_{i=1}^\infty {w}_i(x;t)\,\hat b_i(t)$ for all times.
Using this, one can compute the action of the field operator $\hat\psi(x)$ on the two class of states,
obtaining:
\begin{eqnarray}
\label{psi-coherent}
\hat{\psi}(x) \ket{\vec{\xi},\vec{\varphi};N,t} = \sqrt{N}
\left( \sum_{j}\sqrt{\xi_{j}}e^{i\varphi_{j}}w_{j}(x,t)\right) \ket{\vec{\xi},\vec{\varphi};N-1,t} \ , \\
\label{psi-Fock}
\hat{\psi}(x) \ket{\vec{k};N,t}= \sum_{j} \sqrt{k_{j}}\, w_{j}(x,t) \ket{k_{1},...,k_{j}-1,...,k_{M};N-1,t}\ .
\end{eqnarray}

\section{Generalized quantum measures}

In order to apply the previous formalism to the theoretical interpretation of the above mentioned
procedure of measuring density profiles, it is useful to recall some
results deduced from the experiment. 

When $M=2$ and the atoms in the lattice just before the release of the
confining potential are prepared in a superfluid state described by (\ref{coherent}),
the picture that is obtained at time $t$ after a free expansion shows a high visibility interference pattern,
with fringe spacing mediated by the wave vector $Q= {md/\hbar t}$,
where $m$ is the atom mass \cite{Andrews}. 
This is to be expected, since in the superfluid phase
essentially all $N$ particles occupy the same quantum state. The roughness and imperfection
of the interference figure in a single image, beside to experimental errors, has to be ascribed
to the finiteness of the particle number $N$. 
Indeed, in many-body physics, one can assimilate ensemble averages with
mean values with respect to macroscopically occupied many-body states, provided the number of particles involved
is large enough. Therefore, the larger the number $N$ of atoms the system contains, 
the better a {\it single} absorption image will model the average 
$n_{\vec{\xi},\vec{\varphi}}(x,t)$ of the density
operator at point $x$, 
\begin{equation}
\label{density}
\hat n(x)\equiv \hat\psi^\dagger(x)\,\hat\psi(x)\ ,
\end{equation}
in the state $\ket{\vec{\xi},\vec{\varphi};N,t}$. Using (\ref{psi-coherent}), one can easily compute
the theoretically predicted density profile to obtain
\begin{eqnarray}
\nonumber
n_{\vec{\xi},\vec{\varphi}}(x,t) &\equiv& \bra{\vec{\xi},\vec{\varphi};N,t} 
\hat{n}(x) \ket{\vec{\xi},\vec{\varphi};N,t}
  = N \bigl| \sum_{j}\sqrt{\xi_j}e^{i\varphi_j}w_j(x,t) \bigr|^{2} \\
\label{n-coherent}
&=& N|w(x,t)|^2\Big\{ 1+2\sum_{j<l}\sqrt{\xi_j\xi_l} \cos(\theta_{jl}(x,t)-\varphi_j+\varphi_l)\Big\} \ ,
\end{eqnarray}
where, recalling (\ref{evolw}), the relative dynamical phases take the explicit form 
$\theta_{jl}(x,t)=$\break $\frac{m}{2\hbar t}[-(x-x_j)^2+(x-x_l)^2]$; 
since for the lattice site positions one has $x_j=jd$, it can also be rewritten as 
$\theta_{jl}(x,t)=Q(j-l)(x-\frac{d}{2}(j+l))$, thus reproducing (for $M=2$) the observed
interference pattern mediated by the wave vector $Q= {md/\hbar t}$.

Since taking the picture of the expanded gas is a destructive operation (the sample is lost),
one usually repeats many times the whole measuring process, preparing the system in the
same superfluid initial state. In the case $M=2$, all subsequent absorption pictures
show the same interference figure (modulo experimental uncertainties),
with the same fringe spatial positions. Therefore, by superimposing all these
pictures, one obtains an average image that is indistinguishable from any of the single shot
pictures, and therefore it is again described by the pattern (\ref{n-coherent}).

If instead one starts at $t=\,0$ with a system prepared in an incoherent state, like the
Fock states in (\ref{Fock}), with no definite relative phase relations, 
no interference pattern in the absorption images is expected to appear. 
Surprisingly, this is not the case: in any experiment so far performed,
single shot images always show interference figures compatible with the pattern
in (\ref{n-coherent}), irrespectively from the state of the sample at the moment
of trap release. However, unless the sample is prepared in a superfluid state,
the absolute spatial position of fringes in repeated single shot images is seen to vary
randomly; as a consequence, by superimposing many single shot images,
the interference pattern indeed completely disappears. 

The usual theoretical interpretations of these experimental facts is based on the assumption that
the superposition of single shot images reproduces operator averages. More precisely,
the procedure of repeatedly preparing the system at $t=\,0$ in a generic many body state 
described by the density matrix $\rho$,
and then superimposing all the single shot pictures
taken after a free evolution of the system up to time $t$,
$\rho\mapsto \rho(t)$, is commonly believed to reproduce the quantum average
\begin{equation}
\label{trace}
n_{\rho}(x,t)={\rm Tr}\big[\hat n(x)\, \rho(t) \big]\ .
\end{equation}
For instance, for generic $M$, starting from a Fock state (\ref{Fock}), $\rho=\ket{\vec{k};N}\bra{\vec{k};N}$, 
in this way one would obtain
\begin{equation}
\label{n-Fock}
n_{\vec{k}}(x,\tau) \equiv \bra{\vec{k};N,t} \hat{n}(x) \ket{\vec{k};N,t}
 = \sum_{j=1}^{M} k_{j} |w_j(x,t)|^{2}\ ,
\end{equation}
which shows no interference figures.

An alternate, different theoretical formalization of the experimentally used measurement technique
is nevertheless possible. It takes into account the previously mentioned result
that, for $M=2$, single shot images always present interference patterns
compatible with the profile in (\ref{n-coherent}), irrespectively from the initial state $\rho$,
and extends it to the case $M>2$.
In other terms, the procedure of ``taking a photograph'' of the
expanded sample seems to select a coherent-like state 
$\ket{\vec{\xi},\vec{\varphi};N,t}$ for the state of the system, with definite amplitudes $\xi_i$
and phases $\varphi_i$, and as a consequence produce
the profile $n_{\vec{\xi},\vec{\varphi}}(x,t)$ in (\ref{n-coherent}) for the average density.%
\footnote{From a different perspective, 
this empirical fact has recently been the object of various investigations \cite{Stringari}-\cite{Demler}.}
As seen in experiments, amplitudes and phases nevertheless vary from shot to shot.%
\footnote{Unless, as mentioned before, one starts at $t=\,0$ 
already with a coherent state $\ket{\vec{\xi},\vec{\varphi};N}$,
{\it i.e.} in a superfluid phase.}
The occurrence of given values $\vec\xi$ and $\vec\varphi$ for such parameters
in a single shot will be determined by the initial state $\rho$.
More precisely, the distribution of $\xi_i$ and $\varphi_i$
over many absorption pictures will be determined by the probability
$\bra{\vec{\xi},\vec{\varphi};N,t} \rho(t) \ket{\vec{\xi},\vec{\varphi};N,t}$, which
gives the weight of the configuration $\ket{\vec{\xi},\vec{\varphi};N,t}\bra{\vec{\xi},\vec{\varphi};N,t}$ 
in the expansion of $\rho(t)$ in the basis of coherent states.

As a result, in this scheme, the mean value of the density $\hat n(x)$, as for any other observable,
is given by the sum over all possible configurations $\{\vec\xi,\vec\varphi\}$ of the
average $n_{\vec{\xi},\vec{\varphi}}(x,t)$ weighted with the above mentioned probability;
explicitly, one then should write:
\begin{equation}
\tilde n_{\rho}(x,t) = \int d\mu(\varphi)\ \int d\mu(\xi)\ 
\bra{\vec{\xi},\vec{\varphi};N,t} \rho(t) \ket{\vec{\xi},\vec{\varphi};N,t}\
n_{\vec{\xi},\vec{\varphi}}(x,t)\ ,
\label{n-povm}
\end{equation}
where for the the properly normalized volume and integration measure, 
we have used the shorthand notation [{\it cf.} (\ref{overcomplete})]:
\begin{equation}
\int d\mu(\varphi)\ \int d\mu(\xi)\equiv
\dfrac{(N+M-1)!}{N!} \int_{0}^{2\pi}\frac{d\varphi_{1}}{2\pi}...\frac{{\ud\varphi_{M-1}}}{2\pi} \int_{0}^{1} 
\ud\xi_{1} ... \int_{0}^{1-\xi_{1}-\xi_{2}...-\xi_{M-2}} \ud\xi_{M-1}\ .
\label{measure}
\end{equation}

The expression (\ref{n-povm}) for the averaged density corresponds to a quantum mechanical
generalized measure. It can be described by an operation of trace of $\hat n(x)$ over
the density matrix $\tilde\rho$ obtained from the starting state $\rho$
through the action of a (completely positive) map; explicitly:
\begin{equation}
\label{tilderho}
\rho(t)\mapsto\tilde\rho(t)=\int d\mu(\varphi)\ \int d\mu(\xi)\ V(\vec\xi,\vec\varphi;N,t)\
\rho(t)\ V(\vec\xi,\vec\varphi;N,t)\ ,
\end{equation}
with
\begin{equation}
V(\vec\xi,\vec\varphi;N,t)=\ket{\vec{\xi},\vec{\varphi};N,t}\bra{\vec{\xi},\vec{\varphi};N,t}\ ;
\end{equation}
it is a realization of a {\it Positive Operator Valued Measure} (POVM) \cite{Takesaki}-\cite{Alicki}.

Notice that the mean value for the atom density given in (\ref{n-povm}) obtained through the POVM,
{\it i.e.}
\begin{equation}
\label{mean-povm}
\tilde n_{\rho}(x,t) = {\rm Tr}\Big[\hat n(x)\, \tilde\rho(t)\Big]\ ,
\end{equation}
is clearly distinct from that obtained using the definition (\ref{trace});
nevertheless, they both reproduce the profile (\ref{n-coherent}) in the case
of a superfluid initial state, $\rho=\ket{\vec{\xi},\vec{\varphi};N}\bra{\vec{\xi},\vec{\varphi};N}$,
thanks to the (large $N$) orthogonality of the coherent states. On the other hand,
in the case of a Fock state, $\rho_{\vec k}=\ket{\vec{k};N}\bra{\vec{k};N}$, the explicit evaluation
of (\ref{mean-povm}) gives
\begin{equation}
\tilde n_{\vec k}(x)=\frac{N}{N+M}\sum_{j=1}^{M} (k_{j}+1) |w_j(x,t)|^{2}\ ,
\end{equation}
which differs from the result (\ref{n-Fock}) obtained applying (\ref{trace}) by an overall
normalization factor, and in the weight assigned to the contribution of every single lattice site.
Although in line of principle these differences might have experimental relevance, their
actual detection looks technically problematic, except perhaps in the case of a double-well
potential ($M=2$) \cite{ABFT}. Instead, as we will see in the next Section, density correlations
appear more suitable for studying the predictions of the POVM prescription.

\section{Correlation functions}

Besides for density estimations, absorption images can also be used to extract
information on density correlations: by analyzing the absorption figure in distinct
points one can study correlations in atom positions. Let us then introduce the two-point
correlation function as the average of the following two-point operator%
\footnote{It differs from the density-density correlation operator by a $\delta(x-x')$ contribution; on average,
this term is suppressed by a factor $1/N$ and therefore negligible in the large $N$ limit \cite{Bloch2}.}
\begin{equation}
\label{correlation}
\hat{n}(x,x') \equiv \hat{\psi}^{\dagger}(x)\hat{\psi}^{\dagger}(x')\hat{\psi}(x)\hat{\psi}(x')\ ,
\end{equation}
that, according to the two measuring interpretations discussed in the previous Section, can be
either computed using the analog of the standard trace formula (\ref{trace}),
\begin{equation}
n_\rho(x,x',t)={\rm Tr}\big[\hat n(x,x')\ \rho(t)\big]\ ,
\end{equation}
or by assuming that a single shot picture effectively projects the system state into
a coherent one, so that operator averages should be taken using the transformed
density matrix $\tilde\rho(t)$ of (\ref{tilderho}), thus giving
\begin{equation}
\tilde n_\rho(x,x',t)={\rm Tr}\big[\hat n(x,x')\ \tilde\rho(t)\big]\ .
\end{equation}

In order to make an easier comparison with
the actual experimental data, a further integration with respect to the barycenter
coordinate, $R=(x+x')/2$, is usually performed, 
so that the resulting integrated correlation function depends only on the
relative coordinate $r=x'-x$ \cite{Bloch2}. Then, with a suitable normalization, one is lead to study
the behaviour of the following two functions
\begin{equation}
\label{corr}
\mathcal{G}_{\rho}(r,t) \equiv \dfrac{\int \ud R \ {n_{\rho}(R-\frac{r}{2},R+\frac{r}{2},t)}}
{\int \ud R \ {n_{\rho}(R-\frac{r}{2},t)} \ {n_{\rho}(R+\frac{r}{2},t)}} \ ,
\end{equation}
and, by replacing the standard averages with the POVM ones,
\begin{equation}
\label{corr-povm}
\tilde{\mathcal{G}}_{\rho}(r,t) \equiv \dfrac{\int \ud R \ {\tilde n_{\rho}(R-\frac{r}{2},R+\frac{r}{2},t)}}
{\int \ud R \ {\tilde n_{\rho}(R-\frac{r}{2},t)} \ {\tilde n_{\rho}(R+\frac{r}{2},t)}} \ .
\end{equation}
These functions measure the conditional probability of finding two atoms in points separated by a distance $r$,
averaged over all positions. In absence of correlations, they take a constant value equal to one, while values
greater than one would signal the tendency of the atoms to aggregate, a typical behaviour for bosons.
As in the case of the average density discussed in the previous Section, 
the expressions (\ref{corr}) and (\ref{corr-povm}) take a particularly simple and compact form when the system
is initially prepared in a Fock state, $\rho=\ket{\vec{k};N}\bra{\vec{k};N}$.
Indeed, recalling (\ref{evolw}), that gives the free evolution of the Wannier functions,
one explicitly finds
\begin{equation}
\label{G}
\mathcal{G}_{\rho}(r,t)=\dfrac{N(N-1)}{N^2}\bigg\{ 1 \ + \ \dfrac{1}{N(N-1)}
\sum_{i\neq j=1}^M k_i k_j e^{iQ(i-j)r}\bigg\} \ ,
\end{equation}
while
\begin{equation}
\label{G-povm}
\tilde{\mathcal{G}}_{\rho}(r,t)=\dfrac{N(N-1)}{N^2}\bigg\{ 1 \ + \ \dfrac{1}{(N+M)(N+M-1)}
\sum_{i\neq j=1}^M (k_i+1)(k_j+1) e^{iQ(i-j)r}\bigg\} \ ,
\end{equation}
where, as before, $k_i$ represents the initial occupation number of the $i$-th lattice site and $Q=md/\hbar t$.
It turns out that for a given initial state, the overall profile of (\ref{G}) and (\ref{G-povm}) is very similar:
both functions are essentially constant and close to one almost everywhere, except at the origin
and at positions multiple of $2\pi/Q$ where sharp peaks occur. 

Nevertheless, the two expressions do
differ in the normalization of the oscillator terms, as well as in their dependence on the occupation
numbers. One can check that the effects of these differences become more and more visible as the
initial configuration differs from that described by a balanced Fock state, 
the one with an equal number $N/M$ of atoms in each lattice site. Indeed, for such a state,
up to terms of order $1/N$, one finds that both functions (\ref{G}) and (\ref{G-povm}) reduce to:
\begin{eqnarray}
\label{comodo10}
\mathcal{G}(r,t) &=& 1 \ + \ \dfrac{\sin^2(\pi NQr)}{N^2 \ \sin^2(\pi Qr)} \\
&\simeq& 1 \ + \ \sum_{j=-\infty}^{+\infty} \delta\left( \frac{Qr}{2\pi}-j\right)  \ .
\end{eqnarray}

In order to appreciate the differences between the two theoretical predictions $\mathcal{G}_{\rho}(r,t)$ and
$\tilde{\mathcal{G}}_{\rho}(r,t)$, one has therefore to prepare the system in an unbalanced
Fock state. The way in which this can be experimentally realized is by superimposing a second weaker periodic
potential to the original one; this is done through the
introduction of a second laser directed along the optical lattice, whose wavevector $\kappa_2$
is chosen to be incommensurate with respect to $\kappa_1=\pi/d$, the wave vector
of the first, original one \cite{Fallani1}-\cite{Fallani4}. When the amplitude of the second laser
is weak, thus introducing only small perturbations to the original potential, the dynamics
of the atoms in the lattice can still be described by the Bose-Hubbard Hamiltonian (\ref{BH}),
where however the site depth energy $\epsilon_i$ gets a further site dependent contribution
of order $V_2\, \sin^2(i\kappa_2\pi/\kappa_1)$, proportional to the strength $V_2$ of the
second laser. 

Then, in order to prepare the system in a Fock state, one gradually increases 
the intensity of the lasers, so that the hopping term in the Hamiltonian
(\ref{BH}) becomes negligible. In this way, one drives the system into a ground state
of the form $\ket{\vec{k};N}$, characterized by the filling configuration $\vec k$,
minimizing the total energy $E=\sum_j \epsilon_j k_j + \frac{U}{2} \sum_j k_j(k_j-1)$
with the constraint $\sum_j k_j=N$, for $k_j$ integer. The advantage of working with
a bichromatic lattice is now apparent: it allows to change the distribution of
the $N$ atoms in the $M$ wells by varying the amplitude $V_2$ of the second
potential.

For a given value of the characteristic, physical parameters entering the Hamiltonian (\ref{BH}),
the configuration $\vec k$ that minimizes the energy can be efficiently obtained using a numerical simulation. 
We have used a Monte Carlo method
implementing a {\it simulated annealing} algorithm, that better conforms to the
experimentally adopted procedure, since it considers all
atoms in the lattice at once.%
\footnote{This simulation procedure should be contrasted with the one adopted in \cite{Fallani3},
where the $N$ atoms are inserted in the lattice one by one, while minimizing the energy at
each step.}
Indeed, in a typical experiment, the atoms are first cooled
in an harmonic potential, at the center of which the optical lattice is then slowly
raised. Initially, the hopping dynamics allows a redistribution of the atoms in the various sites;
however, this becomes highly suppressed at regime, leaving the system in a Fock ground state.

To be as closer as possible to an actual experimental situation, we have chosen to work
with the physical parameters that define the apparatus described in Ref.\cite{Fallani2}. Although
the principal lattice is in that case three-dimensional, the second, weaker potential
is switched on only along one direction, making the whole system effectively behaving as a 
collection of independent, separate one-dimensional bichromatic lattices. The whole
system is filled with about $3\times 10^5$ particles, but any single one-dimensional
lattice is formed by about $M=130$ sites, filled with roughly $N=170$ ultracold atoms.

The result of the simulation is summarized in Figure 1:
it shows the plots of the corresponding correlation
functions $\mathcal{G}_{\rho}(r,t)$ (green line) 
and $\tilde{\mathcal{G}}_{\rho}(r,t)$ (red line). 
As expected, these functions are almost everywhere equal to one, except for the presence of
periodic bumps. The difference between the two is particularly visible
in the lower, secondary peaks, whose height is highly suppressed when the
correlations are computed using the POVM prescription.
This difference become more and more evident as the strength of laser giving rise to the
second potential is increased, as clearly shown in Figure 2. As a result, the suppression of the secondary peaks
can be made visible well beyond any statistical error, thus becoming experimentally relevant.

\begin{figure}[h!]
\begin{center}
\rotatebox{-90}{
\includegraphics[scale=.5]{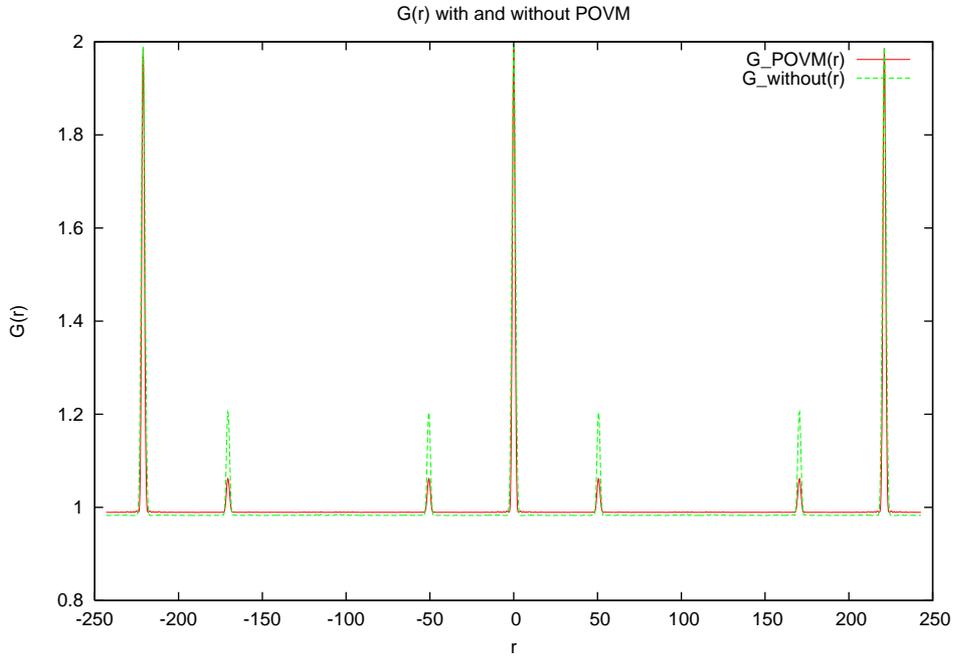}
}
\caption{{\small Behavior of $\mathcal{G}(r,t)$ (green line) and of $\tilde{\mathcal{G}}(r,t)$ (red line) with
$N=170,M=130,V_2=h \times 9.9 \text{ kHz}$.}}
\end{center}
\end{figure}

\begin{figure}[h!]
\begin{center}
\rotatebox{-90}{
\includegraphics[scale=.5]{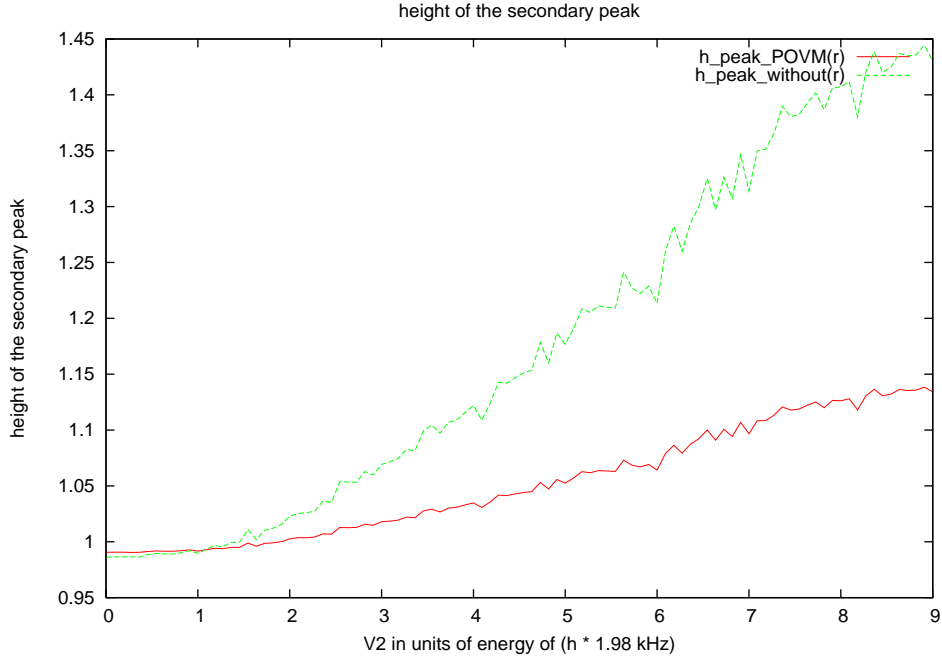}
}
\caption{{\small Height of the secondary peaks as a function of the intensity of the secondary laser
with (red line) and without POVM (green line).}}
\end{center}
\end{figure}

\section{Outlook}

The measuring procedure commonly used in experiments with ultracold gases,
consisting in extracting density profiles from absorption images
taken after the release of the optical lattice, suggests a theoretical interpretation in terms of
generalized quantum measurement processes. The experimental evidence regarding the presence of interference fringes
in single shot absorption pictures irrespective from the initial state of the system suggests the use
of a POVM based on coherent-like, fixed phase states. Within this framework, the averages of system observables
in general differ from those obtained through mean values of the corresponding operator in the system state,
the usually adopted paradigm in interpreting experimental data.

As discussed in Section 4, these differences are hardly visible in the case of density
measures, since they are of the order of the inverse total number of atoms in the sample.
Instead, the situation appears quite different for density-density correlations.
As explicitly discussed in the previous Section, by preparing the system in a suitable Fock
state, the profile of the integrated and normalized correlation function is characterized
by a double series of peaks, the smaller of which appear to be much lowered when the correlations
are computed using the POVM prescription. This effect is very pronounced and surely well
beyond statistical errors for situations close to the actual experiment: 
this may open the way to a direct test of the POVM assumption.

\section*{Acknowledgements}
This work is supported by the MIUR project
``Quantum Noise in Mesoscopic Systems''.

\end{document}